\documentstyle{lamuphys}
\makeatletter
\let\chapter\hid@chapter
\makeatother
\begin{document}
\pagenumbering{arabic}
\title{Distant Supernovae and Cosmic Deceleration
\thanks{To appear
in The Early Universe with the VLT, ed. J. Bergeron, (Berlin: Springer)}
}

\author{Bruno Leibundgut and Jason Spyromilio}

\institute{European Southern Observatory\\
Karl-Schwarzschild-Strasse 2\\
D-85748 Garching\\ Germany}

\maketitle

\begin{abstract}
Distant supernovae can now be detected routinely.
To date 34 supernovae at $z > 0.1$ have been discovered. 
Among them are 12 Type~Ia supernovae confirmed
spectroscopically and suited to measure the cosmic deceleration when 
appropriately employed as standard candles. 
However, peak magnitudes have been determined for only two objects so far
and a determination of $q_0$ is not yet possible. 

We describe the current status of the searches and possible
pitfalls of the method which rests on few basic assumptions.
The importance of
sufficient information on the distant events is stressed and the 
observations of SN~1995K are used as an example of the detailed
procedures employed in the analysis. Only spectroscopic classification
and light curves in at least two filter bands provide the basis 
to use correction schemes for the luminosity which have successfully
been established in nearby samples.

Time dilation has been detected acting on the light curve of SN~1995K
at a redshift of 0.478, providing clear evidence of universal
expansion. The observations are fully consistent with local Type Ia
supernovae in an expanding universe but 
incompatible with the expectations from a static universe.

The contributions of the new, large telescopes to this research 
area are described.
The extension of the observations to even more distant objects will
provide a better leverage to distinguish between the possible
decelerations and the inclusion of Type II supernovae into the sample
add an independent check on the cosmological distances.
\end{abstract}

\section{Introduction}

Observational cosmology has made use of supernovae for several decades now
(Zwicky 1965, Kowal 1968).
Their large luminosity makes them suitable for detection at
the distances required for
the measurement of cosmological parameters. The main drawback of supernovae
 is the brief time they
shine at this luminosity and that they are intrinsically very rare events.

In principle, all supernova types can be used for cosmological purposes.
 Explosions of massive stars,
which are normally recognized by strong hydrogen lines in their
spectrum (Type II supernovae), can be physically
 understood when the radiation
hydrodynamics are solved (Eastman et al. 1996) and yield distances through
 the expanding photosphere
method (Schmidt et al. 1992, 1994). The explosive carbon incineration of
 white dwarfs produces very
luminous events with no hydrogen in their spectra - Type Ia supernovae
(SNe~Ia).
 These objects display a fairly
uniform behavior which led to their use as standard candles (Kowal 1968,
Tammann \& Leibundgut 1990, Branch \& Miller 1993,
Sandage et al. 1996). The simple adoption of 
a unique peak luminosity, however, has recently
been challenged by SN Ia samples with accurate relative distances 
and a dispersion of 0.6 magnitudes in maximum
luminosity (Phillips 1993, Hamuy et al. 1995). The
uniformity of the class is further eroded by correlations found in the light
curve shapes and spectral
appearances (Phillips 1993, Hamuy et al. 1995, Nugent et
al. 1995, Vacca \& Leibundgut 1996). 
Local samples can be corrected by the proposed light curve 
shape fitting and only a small residual scatter ($<$0.2 mag.) remains
(Hamuy et al. 1995, Riess et al. 1996).

Despite the variety of SNe~Ia in the nearby universe we are still
confident that the mechanisms are in place to use them at large
distances for the determination of cosmological parameters, provided
sufficient observations of each distant event are available.

The value of the Hubble constant has been the main focus of  SN Ia
applications (Sandage et al. 1996, Branch et al. 1996,
Hamuy et al. 1995, Riess et al. 1996). 
The discrepancies in the reported results can be traced to the
 absolute nature of the measurement.
The discussion of Riess et al. (1996) exemplifies the positions and
 assumptions adopted by several groups
and resolves some of the conflicts rather convincingly.

Measuring the cosmic deceleration is fundamentally simpler, since it is done
 through relative measures. It is based on the comparison of the
 apparent brightness of objects with known relative luminosity at largely
 different redshifts.
The ability to find the luminosity of a distant SN~Ia relative to a nearby
 twin rests purely on
observables (Schmidt et al. 1996, Perlmutter et al. 1996).

In this paper we present the basic ingredients necessary for successful
 detection and observation of distant
supernovae and describe the current status of
 projects which use SNe Ia to
measure $q_0$ (\S 2). 
In section 3 we develop the time dilation test for universal
 expansion as applied to SN 1995K.
The prospects for and the contributions of the VLT
 to experiments like this one are presented together with the
 conclusions.

\section{Observing distant supernovae}

Finding distant ($z>0.3$) supernovae is hampered by their variability
 and the rareness with which
they occur. To overcome these obstacles a massive approach is needed. This has
 become possible with
the development of  sensitive, panoramic CCD detectors. 
Faint brightness levels have to be reached 
to increase the period a supernova is detectable. 
At the same time a large number of galaxies has to be
surveyed to increase the chance of catching a supernova in the act. 

There are a few pitfalls in the use of  supernovae for accurate cosmological
 distance determinations.  The
first one - to find the objects - has been overcome by the concerted efforts
 at 4m telescopes to image many
galaxies in a short time period. Typically prime focus cameras with their
 large field of view are employed.
The search fields, some including known clusters at z$>$0.3, are observed 
after new moon to be re-imaged at the
beginning of the next dark period. This ensures that all events detected
 during the second run are still on
their rising branch and their light curve peak can be observed
(Perlmutter et al. 1996). The currently
employed strategy
detects several supernovae (5--10) in such a search period (Garnavich et al.
1996a,b, Perlmutter et al. 1995b, 1996). 
All candidate objects for which it was possible to obtain a spectrum 
have been confirmed as supernovae at large distances.

Spectroscopic classification is of paramount importance as only Type Ia
 supernovae should be used for this
method. Furthermore, on has to discriminate against SNe~Ia which show clear
peculiarities (SN 1991bg: Filippenko
et al. 1992a, Leibundgut et al. 1993, Turatto et al. 1996; SN 1992K: Hamuy et
al. 1994; SN 1991T:
Filippenko et al. 1992b, Phillips et al. 1992, Ruiz-Lapuente et al. 1991;
SN 1992bc: Maza et al. 1994) to avoid
contamination of the sample. Since a SN Ia reaches about R=22.5 at z=0.5, 
the required spectroscopy
stretches the capabilities of  the current 4m telescopes to the limit.
The ability to schedule the spectroscopic follow-up observations ahead
of time has proven to be essential for their success.
 Thanks to the combined efforts by
several observatories has it been possible to obtain a spectrum for
a fair fraction of the events (see below).

A further complication is the correction required to compare the observed
 flux evolution to the rest frame
light curves of local supernovae. This K-correction, which depends on the
phase of the supernova, can be overcome in two ways. One possibility is
to use special
filters corresponding to redshifted B and V passbands to minimize 
the corrections and also largely remove the
dependence on phase. The other approach is to use standard broad-band
filters which are closest to B and V in the rest frame 
of the supernova. Such K-corrections for regular SNe Ia have been
calculated (Hamuy et al. 1993, Kim et al. 1996).
In practice, this problem does not contribute significantly to the error
budget.

\begin{figure}
\caption{Comparison of the spectrum of SN 1995K with local SNe Ia near
maximum light. The spectra are plotted in rest frame wavelengths.}
\end{figure}

Finally, photometry in two filters delivers the color of the supernova. 
The multi-color
light curve shape method (Riess et al. 1996) makes use of this information
to correct for possible absorption. Potentially absorption is one of the
major contributors to the uncertainty in the derivation of $q_0$. 

The first distant supernova was discovered in a dedicated search already
in 1988. The object located in a galaxy cluster at z=0.31 (SN~1988U, AC 118) 
was discovered about two weeks after
maximum light and the spectrum suggested a Type Ia classification 
(N\o rgaard-Nielsen et al. 1989). The same search reported a second
event at a somewhat smaller
redshift (SN 1988T, z=0.28; Hansen et al. 1989). The classification of
this object is uncertain.

In recent years two major efforts to find and observe distant supernovae
with the goal to determine the
geometry of the universe have emerged. SN 1992bi was reported at a redshift
of 0.458  by the Berkeley Cosmology Project and proved that it is 
possible to obtain an accurate light curve (Perlmutter et al. 1995a).
The High-Z Supernova Team
(Schmidt et al. 1995, 1996) has reported its observations of SN 1995K which
include light curves in two
filters and a spectrum near maximum light.
The event occurred in a galaxy at a redshift of 0.478 and
appears superposed on a spiral
arm. The galaxy spectrum displays H$\alpha$, H$\beta$, [O~II] 
and [O~III] lines in emission and is consistent
with an Sbc classification. The supernova
 spectrum itself had to be 
corrected for galaxy contamination and heavily smoothed. The result is a
spectrum which displays the
usual absorption and emission features of a Type Ia supernova near maximum
light (Figure 1). All lines are
shifted to the same redshift as determined for the galaxy confirming the
 association. A comparison of the
SN~1995K spectrum with nearby events clearly shows the close relation
to regular SNe~Ia. The closest match is the near-maximum spectrum of SN~1989B.
The comparison with the overluminous SN~1991T
exemplifies the importance of spectroscopic classification. 
All spectral features of SN~1995K are stronger than in SN~1991T.
In particular, the deep absorption due to Si II is characteristic of 
'normal' SNe Ia
\begin{figure}
\caption{B and V rest frame light curves of SN 1995K.
The best $\chi^2$ fits for a simultaneous fit to the dilated (dark) and 
non-dilated (grey) light curves of SN 1991T are shown.}
\end{figure}
(Nugent et al. 1995) and distinguishes SN 1995K from SN 1991T. 
Similarly, the absorption ascribed to
Ti II in low-temperature supernovae as observed in SN 1991bg near maximum 
is missing in SN 1995K. The phase of the spectrum is
independently set by the light curve (see below) and is fully consistent
 with the spectral observation.

Light curves have been obtained in special B and V filters which correspond
 to a redshift of 0.45 and the
Kron-Cousins R and I filters. At the redshift of SN 1995K (z=0.478) 
the transmission curve of the R filter is almost identical with the
redshifted B passband and K-corrections reduce to a constant describing the
zero-point offset between the B and R
passbands (see also Kim et al. 1996). We hence combined the B45 and R data
sets as well as the V45 and
I filter observations, respectively. The photometry 
covers the maximum phase quite adequately (Figure 2). 
The light curves span from about 18 days before until
35 days after maximum in B and from about peak light to 30 days thereafter
in V.
In fact, the first two
measurements are from pre-discovery search observations. 
The supernova was discovered close to maximum light. The detection
was not triggered in the earlier observations due to the
faintness of the object (R$_{\rm max}$$\approx$22.2) and its projection onto
the high surface-brightness disc of the galaxy (e.g. Leibundgut et
al. 1995).
The pre-maximum points are essential for the determination of accurate light
curve fits (Leibundgut et al. 1996) and a reliable measurement of the
peak brightness. 

\begin{figure}
\caption{Histogram of distant supernovae reported before June 1996.}
\end{figure}

As of June 1996 28 supernovae at redshifts larger than 0.3 have been
reported. Of those at least 9 have been confirmed spectroscopically as being of
Type Ia (N\o rgaard-Nielsen et al. 1989, Schmidt et al. 1995, 
Perlmutter et al. 1995b,
Garnavich et al. 1996a, b). Several supernovae without spectra have 
light curves typical of SNe~Ia.
The peak is covered in most cases, but modeling is still required for
an accurate determination of the maximum brightness (cf. Perlmutter et
al. 1996). The histogram of
all supernovae discovered in these campaigns shows a clear
concentration to the range of $0.3<z<0.5$ with 70\% of all events
detected so far (Figure 3).  Monte Carlo simulations indicate that with
the accuracy achieved for SN~1995K it will take about 30 SNe Ia to
decrease the uncertainty on $q_0$ to about 0.1 (1 $\sigma$, Schmidt
et al. 1996). The remaining error is mostly due to the uncertainty in the
normalization by the local supernovae. The
application of corrections like the multi-color light curve shape method
(Riess et al. 1996) require the availability of at least two filter 
light curves and
sufficient photometry coverage. Of course, it also implies that
distant supernovae are indeed relatives of the nearby events and the
training set employed to find the `corrections' according to light
curve shape and color is applicable. 

\begin{figure}
\caption{Hubble diagram of supernovae with the lines of constant
luminosity in different mass universes. The horizontal bars
show the redshifts of
known distant supernovae. Spectroscopically classified SNe~Ia are
indicated by thick bars, SNe~II and unclassified objects by short, thin
lines.}
\end{figure}

The current status of the determination of $q_0$ is summarized in
Figure~4 (cf. also Perlmutter et al. 1996). There are three SNe 
plotted at redshifts larger than 0.3.
Two have spectroscopic SN~Ia classifications (SN 1988U and SN 1995K) and 
the third (SN 1992bi) displays a light curve typical of SNe Ia. SN~1995K
and SN~1992bi have been observed through their maximum and have
fairly secure peak magnitudes, while SN~1988U has been extrapolated to
maximum. Clearly it is too early to estimate the size of any cosmic
deceleration with these three objects. Nevertheless, it is encouraging 
that the two distant objects fall between the lines for $1/2 > q_0 > 0$.
Note that we have used the normalization of the `uncorrected' nearby
sample of Hamuy et al. (1995). No correction for the light curve shape
has been applied to the distant supernovae either.
Light curve shape corrections change the zero-point of the
normalization and have to be applied carefully. This further
exemplifies the importance of a large knowledge base for the distant
supernovae to confirm their conformity with the local samples. 

The redshifts of additional supernovae already discovered in the two projects
are indicated in Fig.~4. It appears that quite a few events have been
recorded to date and that we should be able to find the value
for the deceleration fairly soon. However, not all objects are indeed
SNe~Ia. Some have no spectroscopic confirmation while others are known
to be Type~II. The importance of sufficient light curve coverage 
is another limiting
factor. Figure~4 thus presents a very optimistic view and we will not
be able to fill the diagram with a point for each line indicating a SN
redshift. In the next year a similar amount of distant events
will be discovered and soon we should be able to
measure $q_0$.

\section{Time Dilation}

A significant result, which can be derived from the single event
currently available to us, is the direct observation of time dilation due
to universal expansion. The light curve of a distant supernova
acts as a clock, the ticks of which can be directly compared to
wavefront stretching as observed in the spectra. In an expanding
universe the two have to go together while other theories of redshift
invoke energy dissipation of photons or similar mechanisms (Arp 1987,
Arp et al. 1990). 
Two distant SNe~Ia 
have been observed at least 18 days before maximum (SN 1994am,
Perlmutter et al. 1996, and SN 1995K).
No local event has been observed at such an early phase.
The earliest observations of nearby SNe~Ia are reported 
in the range of 14 days
before peak (e.g. SN~1990N; Leibundgut et al. 1991). Moreover, the brightness
evolution of SN~1995K is the slowest ever observed for any SN~Ia (Leibundgut
et al. 1996).
In Figure 2 we have overplotted the best fits as
determined by minimization of $\chi^2$ to the combined B and V rest
frame dilated and non-dilated light curves. For the comparison we have
chosen one of the slowest local supernovae known, SN~1991T. 
The non-dilated curves cannot fit the observations, while the photometry
of SN~1995K is fully consistent with the dilated template.
Analyses with comparison light
curves of other local events confirm this result (Leibundgut et al.
1996). The combination of regular color, typical spectrum, a
luminosity in the range of reasonable deceleration parameter values,
and the conforming light curve of SN~1995K strongly favor the dilated case. In a
non-expanding universe SN~1995K would represent a weird event
displaying a regular spectrum but deviating strongly from the
locally established luminosity-decline rate relation, and displaying 
the slowest decline of all known SNe~Ia.
Malmquist bias could not explain this
supernova as it would be less luminous than local SNe~Ia in a static
universe. The same conclusions was found
based on light curves in a single filter for a
set of seven distant SNe ($0.3 < z < 0.5$) three of which with a
spectroscopic Ia classification (Goldhaber et al. 1996). 

\section{Conclusions}

Finding supernovae at redshifts above 0.3 has become a routine
enterprise over the last two years. There are currently two groups
vigorously following this route to find the value of the deceleration
parameter (Perlmutter et al. 1995a, 1996, Schmidt et al. 1996,
Leibundgut et al. 1996). Several supernovae are found each semester in
the scheduled search runs. However, the currently available data set is
not quite large enough to make a serious attempt to measure $q_0$. Only
a dozen objects has the required spectral classification and
sufficient light curve coverage to provide the basis for an accurate
measurement of the peak brightness and the correction to the
luminosity from the light curve shape. There also remains too much
slack in the zero-point of the expansion lines in the Hubble diagram
as defined by local supernovae. This uncertainty directly goes into the
estimate of the cosmic deceleration. 

Another error source which has to
be carefully excluded are systematic differences of distant events
from the nearby ones. 
Thus the sample has to be tightly checked against possible
evolutionary effects. The theoretical understanding of SNe~Ia is not
sound enough to exclude evolutionary effects which could change the
luminosities of the explosions (Canal et al. 1996). Not all models
predict such changes but some do. By a careful analysis of the distant
sample we should be able to distinguish among the various current
candidate models of SN Ia explosions.

Sufficient statistics and, eventually,
detailed spectroscopic follow-up observations should be able to detect such
differences.

The large aperture telescopes will provide the necessary spectroscopy
for the classification of the supernovae. Their r\^ole will be
in the detection and follow up observations of more distant events.
SNe~Ia at maximum should be around I$\approx$25 at z=1 and the 
difference between an
empty and a critical universe is 0.54 magnitudes providing more leverage
for the determination of $q_0$. The complication of a cosmological
constant can also be explored with standard candles at large distances
(Goobar \& Perlmutter 1995). The infrared capabilities of
the large telescopes are of paramount importance as the rest frame V light
is shifted into the J band at z=1. 

Important cosmological tests also remain for the VLT and its 8m-class partners.
The time dilation as observed in the spectral evolution of a distant
SN~Ia would provide an even more stringent proof of cosmic expansion.
The distances derived for Type II supernovae through the expanding
photosphere method (Eastman et al. 1996) are based on the combination
of a luminosity distance with an angular diameter distance. These two
distances are related in all cosmological models (Carroll, Press, \&
Turner 1992). Since these explosions are typically less luminous than
SNe Ia only the large telescopes will be able to deliver the signal
needed to measure the expansion velocities accurately.


Acknowledgment: It is a pleasure to thank our collaborators in this
experiment. Some of the data presented in this review has been kindly
provided before publication. 


%
%
%

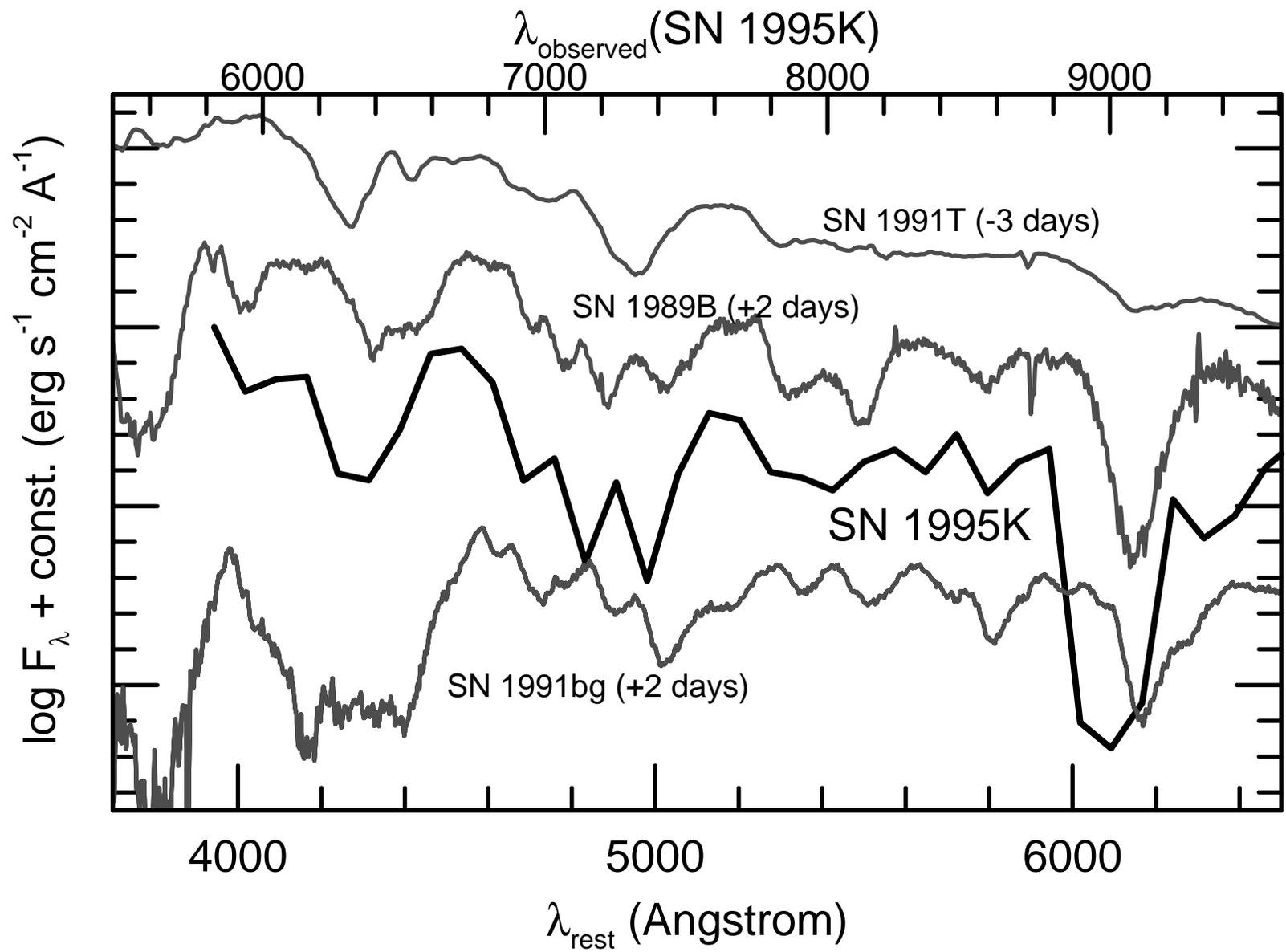
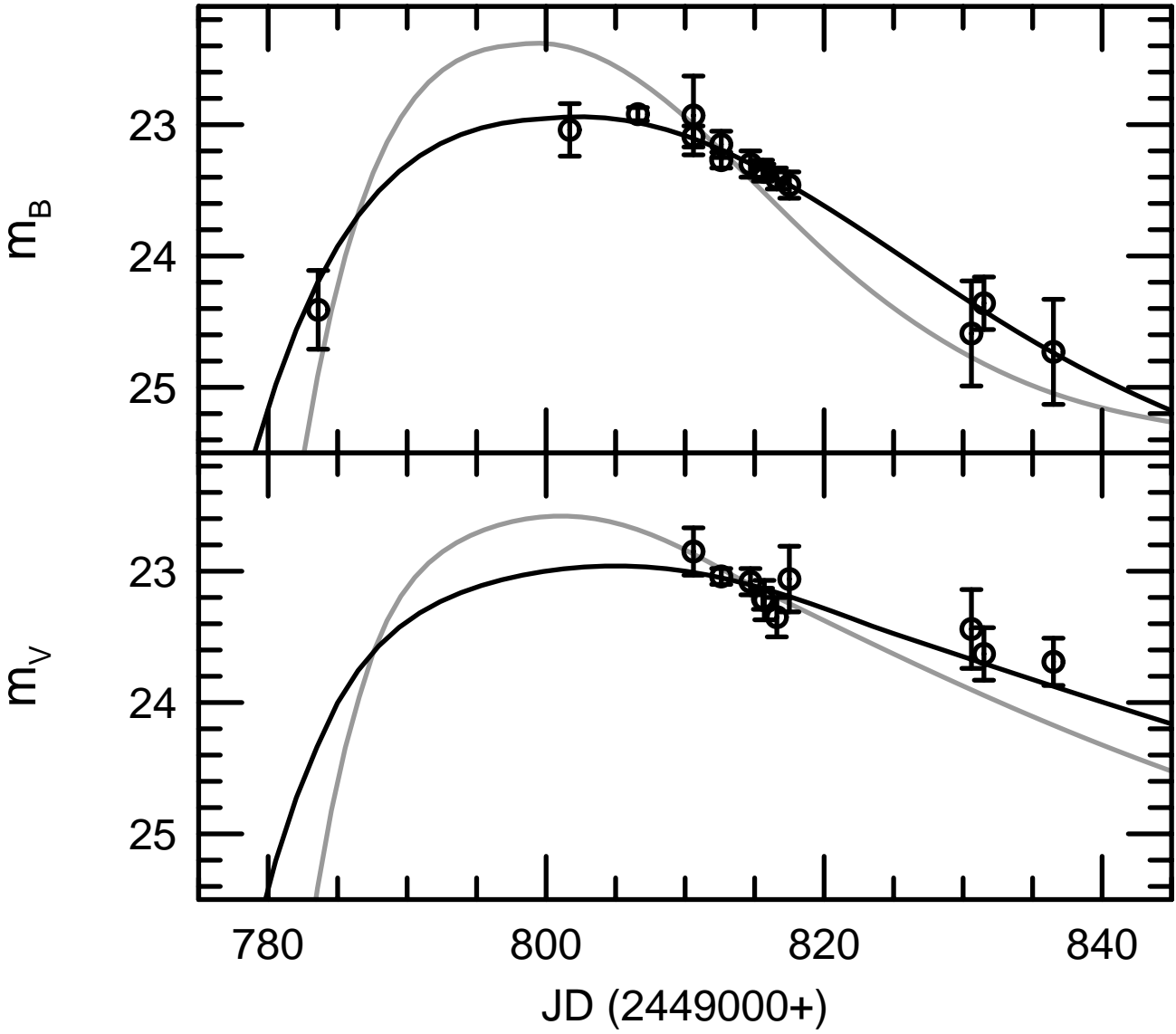
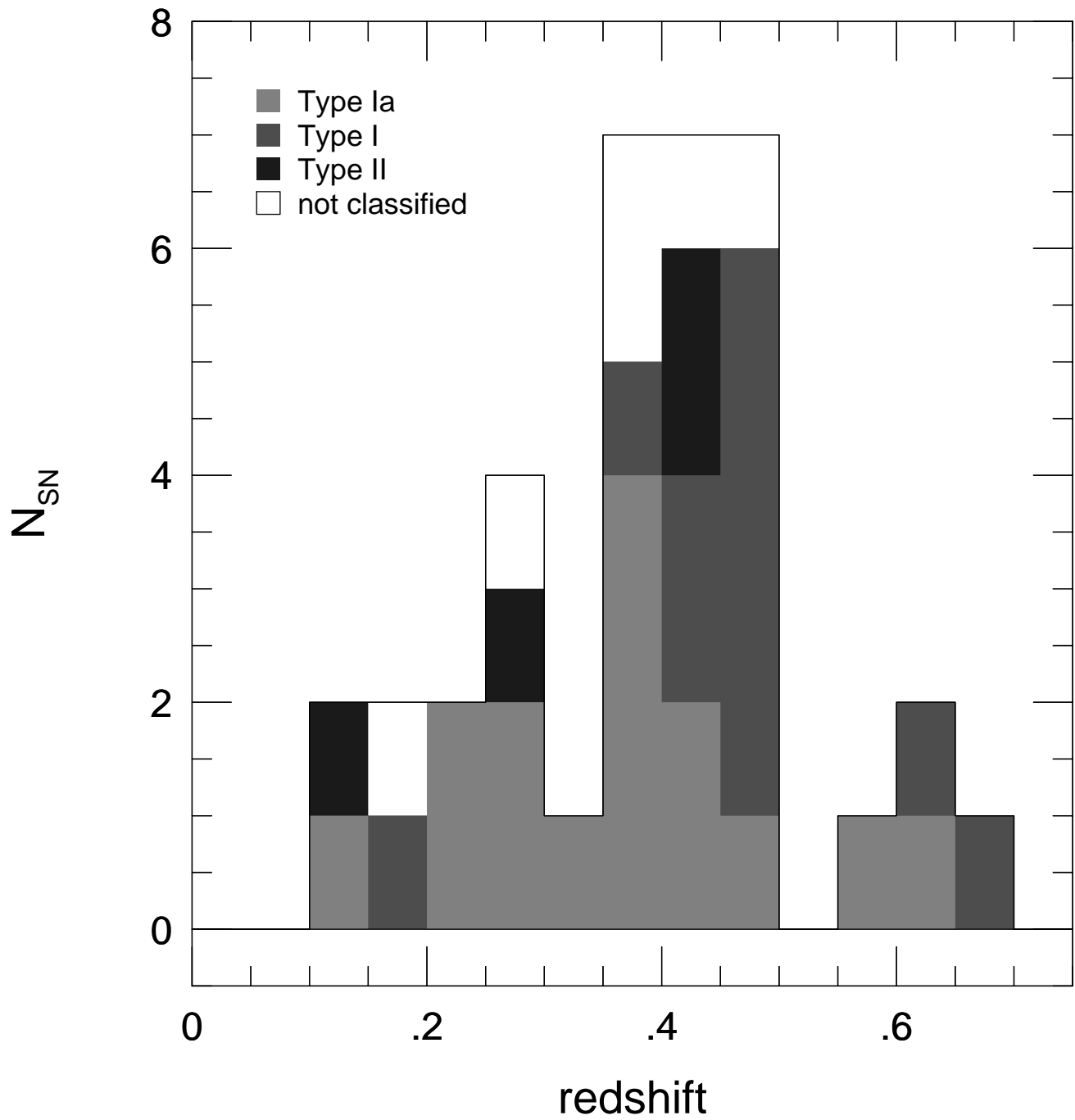
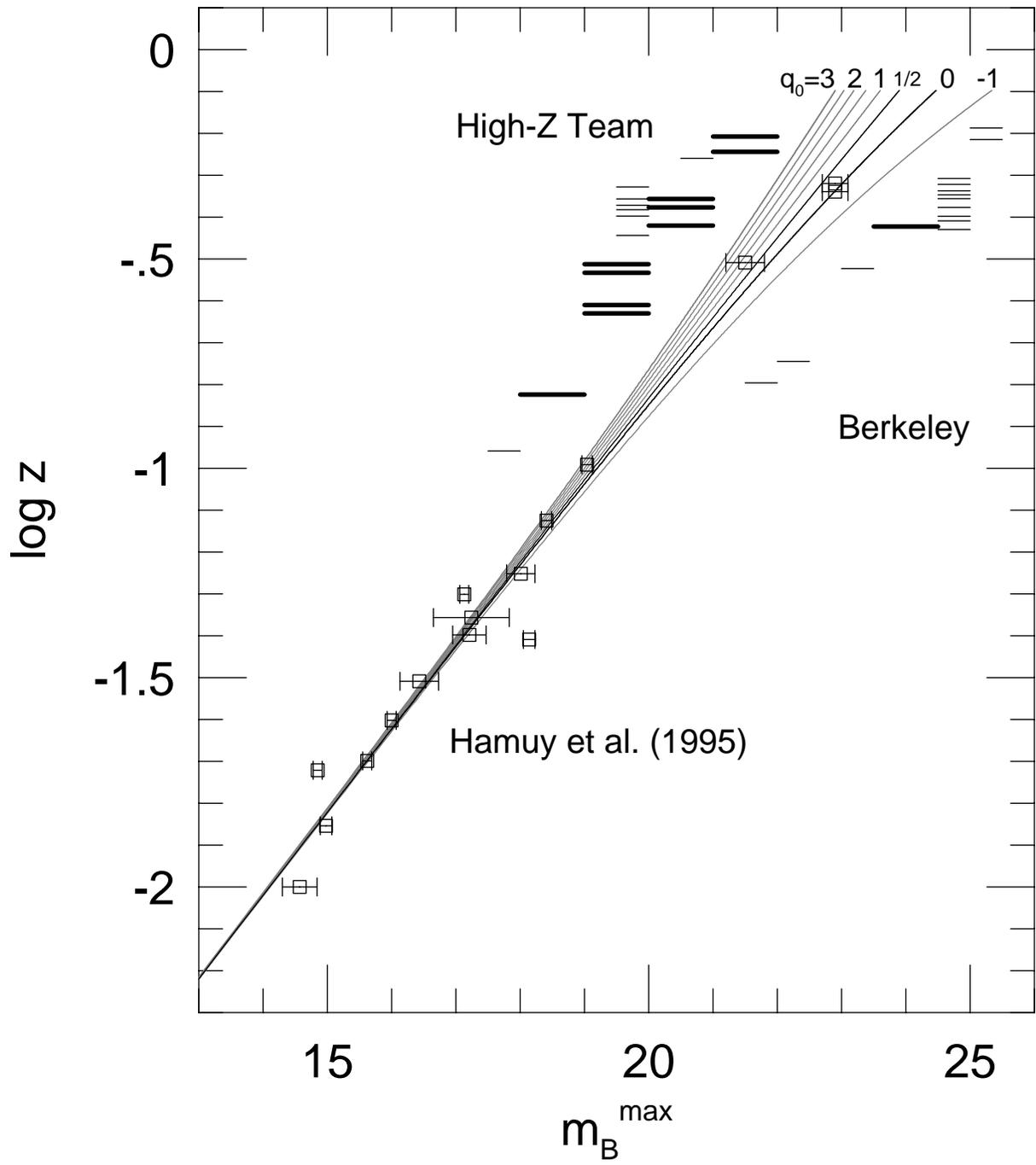
\end{document}